\documentclass[
    ,final           
  ]
  {aipproc}

\layoutstyle{6x9}

\input paperdef

\usepackage{epsfig}
\usepackage{subfigure}
\usepackage{latexsym}
\usepackage{amsmath}
\begin{document}
\begin{flushright}
DCPT/09/158\\
IPPP/09/79\\
\end{flushright}

\vspace{1em}

\title{Precise Predictions for Higgs Production
in Neutralino Decays}


\author{A.C.~Fowler}{
  address={IPPP, University of Durham, Durham DH1~3LE, UK}
}

\begin{abstract}
 Complete one-loop results, supplemented by two-loop Higgs propagator-type corrections,
 are obtained for the class of processes 
$\tilde{\chi}^0_i\rightarrow \tilde{\chi}^0_j h_a$ in the MSSM with CP-violating phases for parameters entering the process beyond lowest order.
The parameter
region of the CPX benchmark scenario where a very light Higgs boson is unexcluded
by present data is analysed in detail. 
We find that the
decay $\tilde{\chi}^0_2\rightarrow \tilde{\chi}^0_1 h_1$ 
may offer good prospects to detect such a light Higgs boson.
\end{abstract}

\maketitle


\section{Introduction}
Higgs physics is one of the main goals of the Large Hadron Collider.  
In the MSSM, the Higgs
spectrum contains five physical
Higgs bosons, the CP-even $h$ and $H$, the CP-odd $A$ and
the two charged $H^\pm$. Higher-order contributions yield large corrections to the masses
and couplings, and, in the complex MSSM, induce CP-violating mixing
between $h,H$ and $A$. If the mixing is such that the coupling of the lightest Higgs boson, $h_1$, to
gauge bosons is significantly suppressed, this state can be very light
without being in conflict with the exclusion bounds from Higgs
searches at LEP and elsewhere~\cite{Barate:2003sz,Schael:2006cr}.  In particular, in the CPX
benchmark scenario~\cite{Carena:2000ks} an unexcluded region remains in which 
$\MHe \approx 45 \gev$ and $\tan\beta \approx 7$~\cite{Schael:2006cr,Williams:2007dc}. Because this region 
will also be difficult to cover at the LHC~\cite{Buescher:2005re,Schumacher:2004da,Accomando:2006ga}, one may need to 
consider additional, non-standard channels. 
Light Higgs bosons can in particular be produced in the decays of 
neutralinos and charginos. 
Noting that higher-order corrections 
can be large in the CPX scenario, we obtain precise predictions for the 
process $\tilde{\chi}^0_i\rightarrow \tilde{\chi}^0_j h_a$, 
where $h_a=h_1,h_2,h_3$.
\section{Renormalisation and Loop corrections}
We have calculated the full one-loop vertex corrections to the process 
$\tilde{\chi}^0_i\rightarrow \tilde{\chi}^0_j h_a$, taking into account the full MSSM contributions and
the phase dependence of the sfermion trilinear coupling, $A_f$, and the gluino mass parameter, $M_3$, 
which are complex in the CPX scenario.  
We assume real bino, wino and higgsino parameters, $M_1$, $M_2$ and $\mu$.
We have made use of the programs \texttt{FeynArts}, 
\texttt{FormCalc}  and \texttt{LoopTools}~\cite{Kublbeck:1990xc,Hahn:2000kx,Hahn:2001rv,Hahn:1998yk},
supplementing the model files with our
counterterms for the vertices involved.

For the neutralino-chargino sector we introduce counterterms of a similar 
form to Ref.~\cite{Fritzsche:2002bi}.  However, we apply different on-shell conditions and allow
 CP-violation, see Ref.~\cite{Fowler:2009ay} for details.
  The field renormalisation constants are fixed by requiring
 diagonal on-shell 1PI two-point vertex functions
and propagators with unity residues.  
For the on-shell parameter renormalisation of $M_1$, $M_2$, $\mu$, we require that 
the loop-corrected pole masses of $\tilde{\chi}^0_{1},\,\tilde{\chi}^0_{2}$ 
and $\tilde{\chi}^{\pm}_{2}$ coincide 
with their tree level values.  

\begin{figure}[tb!]
\centering
\includegraphics[height=2.3cm]{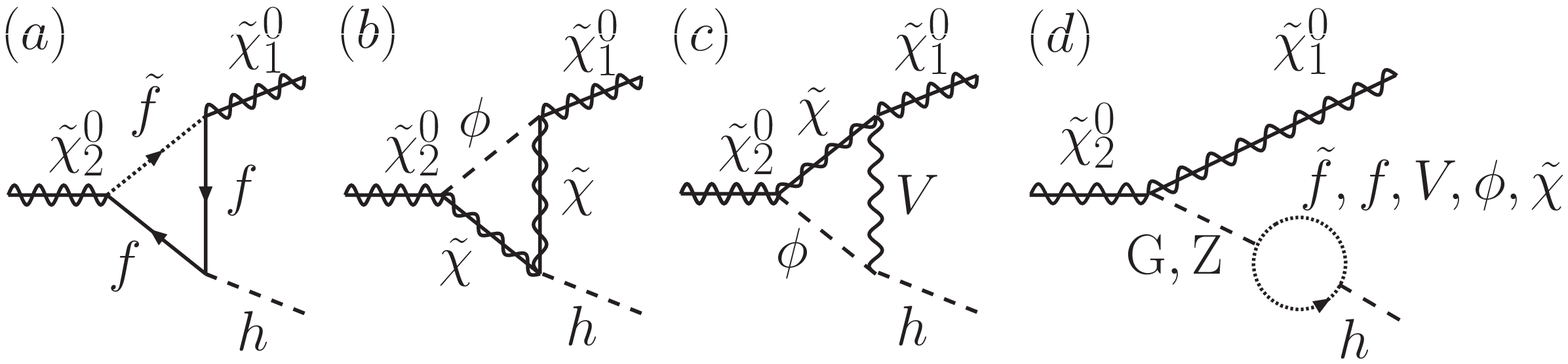}
\caption{Examples of (a-c) 1PI vertex diagrams and (d) reducible G-Z self-energy diagrams.}
\label{diag}
\end{figure}

We renormalise the standard model parameters as in Ref. \cite{Denner:1991kt}, 
while for the Higgs sector we use the combined on-shell/$\overline{\mathrm{DR}}$ renormalisation of
 Ref.~\cite{Frank:2006yh}.  This scheme makes use 
of finite normalisation factors $\hat{\mathbf{Z}}_{ij}$, which automatically include the leading reducible self-energy 
diagrams involving $h,H,A$ beyond the one-loop level.  
 As in Ref.~\cite{Williams:2007dc}, we combine the triangle vertex pieces $\hat{\Gamma}^{\mathrm{1PI}}$ with the reducible pieces  $\hat{\Gamma}^{\mathrm{G,Z.se}}$, such as those in Figs.~\ref{diag}(a-c) and \ref{diag}(d) respectively, 
 as follows,
\begin{equation}
\hat{\Gamma}^{\mathrm{Full\,Loop}}_{\tilde{\chi}^0_{i}\tilde{\chi}^0_{j}
 h_a}=\mathbf{\hat{Z}}_{al}[\hat{\Gamma}^{\mathrm{1PI}}_{\tilde{\chi}^0_{i}\tilde{\chi}^0_{j}
 h_l^0}(M_{h_a}^2)+\hat{\Gamma}^{\mathrm{G,Z.se}}_{\tilde{\chi}^0_{i}\tilde{\chi}^0_{j} h_l^0}(m_{h^0_l}^2)]
\label{eqamp}
\end{equation}
where $m_{h^0_l}$ ($h^0_l=\{h,H,A\}$) and $M_{h_a}$ ($h_a=\{h_1,h_2,h_3\}$) are the tree-level and loop-corrected Higgs boson masses respectively.  

In order to obtain the most precise predictions, 
we have combined our one-loop results with two-loop Higgs propagator-type corrections by using 
two-loop $\mathbf{\hat{Z}}$ factors and Higgs masses $M_{h_a}$ from \texttt{FeynHiggs 2.6.5} \cite{Frank:2006yh,feynhiggs,fhrandproc,mhcpv2l}.  
We compare our loop-corrected results to an Improved Born approximation, which includes only the two-loop Higgs propagator-type corrections.
\section{Numerical Results and Conclusions}
\begin{figure}[htb!]
\centering
\includegraphics[height=5.5cm]{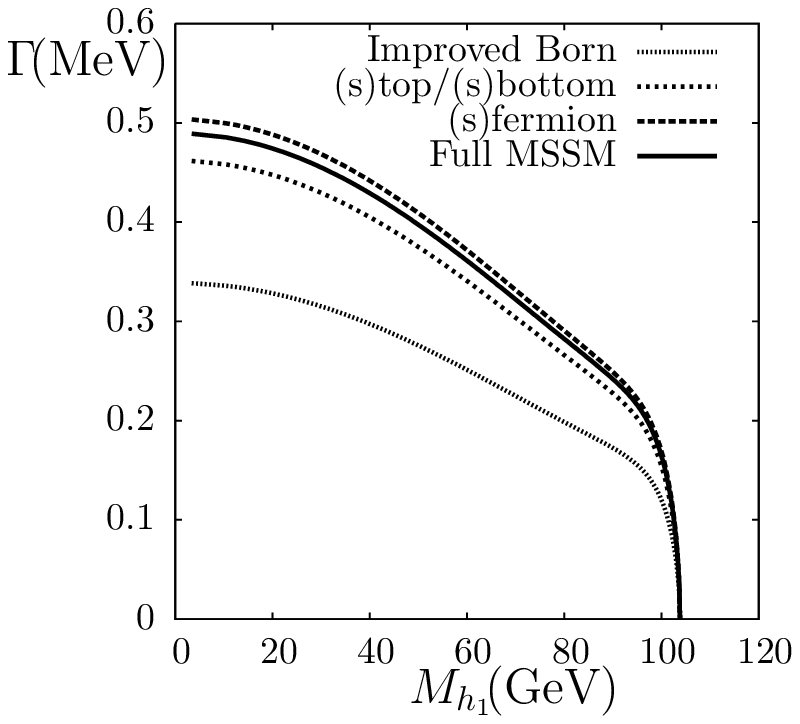}
\includegraphics[height=5.5cm]{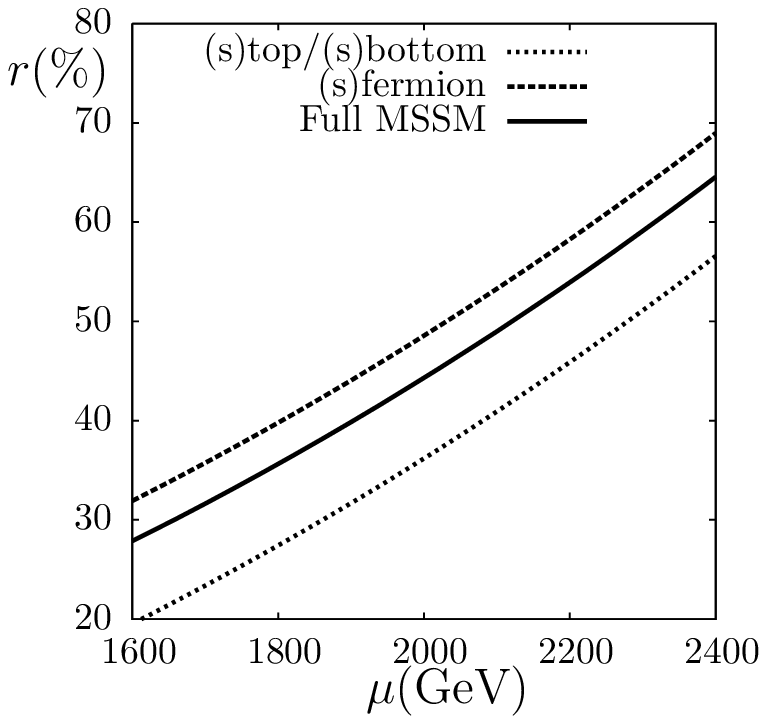}
\label{Mh111}
 \caption{\textit{Left}: Decay width $\Gamma(\tilde{\chi}^0_2\rightarrow\tilde{\chi}^0_1 \mathrm{h}_1)$ plotted against $M_{h_1}$. \textit{Right}: Ratio $r\!\!\!=\!\!\!(\Gamma_{\!\!\mathrm{Full\;Loop}}\!\!\!-\!\!\!\Gamma_{\!\!\mathrm{Improved\,Born}})/\Gamma_{\!\!\mathrm{Improved\,Born}}$ for $\tilde{\chi}^0_2\rightarrow\tilde{\chi}^0_1 \mathrm{h}_1$, plotted against $\mu$ with $M_{h_1}=40\,\mathrm{GeV}$.}
\end{figure}

We now present numerical results 
 for the CPX scenario, using the following parameters unless stated otherwise (in GeV): $\mu=2000$, $M_3=1000 i$, $A_{\mathrm{t,b},\tau}=900 i$, $M_{\mathrm{SUSY}}=500$, $M_2=200$, $M_1=(5/3) \tan{\theta_W}^2 M_2$, $\tan\beta=5.5$ and $m_t=172.4$.  Fig.~\ref{Mh111} (left) shows the partial decay width 
$\Gamma(\tilde{\chi}^0_2\rightarrow\tilde{\chi}^0_1 \mathrm{h}_1)$ as a function of $M_{h_1}$.
 We see that the largest new contribution, adding about $35\%$ onto the Improved Born result, comes from 
the triangle diagrams containing third generation quarks and squarks ($t,\tilde{t},b,\tilde{b}$),
 due to the large top Yukawa coupling.  The other (s)fermions increase the vertex
 contribution to around $50\%$, while the remaining MSSM particles 
reduce the total effect to around $45\%$. 
Such large effects from the genuine vertex corrections are not unexpected in this extreme scenario
 with a large higgsino parameter and trilinear couplings. In Fig.~\ref{Mh111} (right) we show how the percentage effect of our vertex
 corrections increases with $\mu$.
\begin{figure}[htb!]
\centering
\label{argAtratio1}\includegraphics[height=5.5cm]{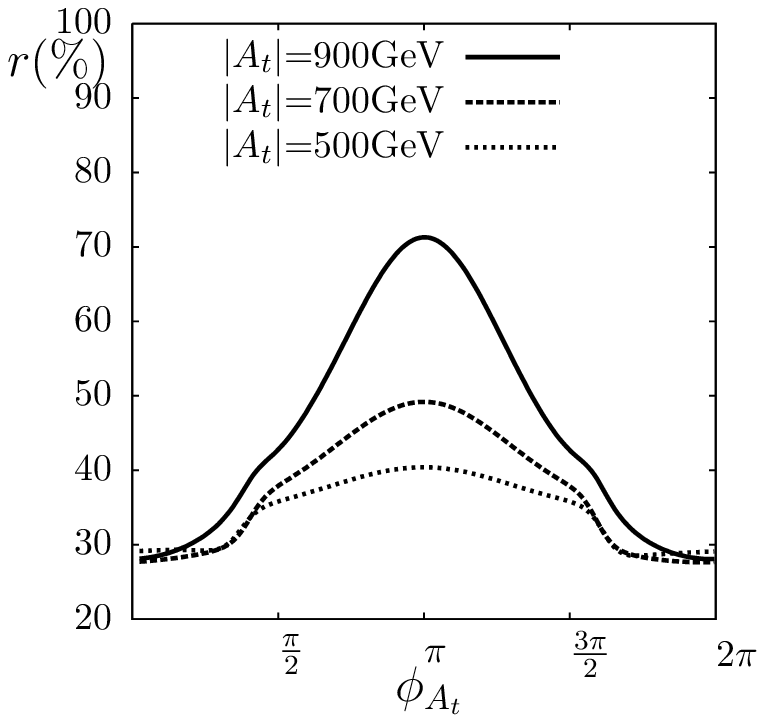}
\includegraphics[height=5.5cm]{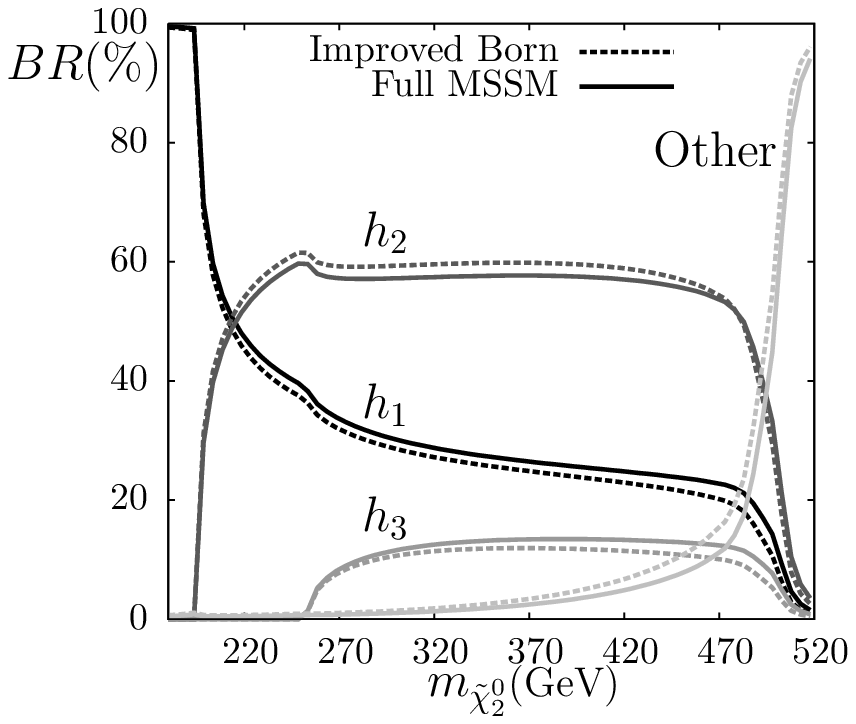}
 \caption{\textit{Left}: $r$ (see Fig.~\ref{Mh111} caption) plotted against $\phi_{A_t}$ for $M_{h_1}=45\,\mathrm{GeV}$, $\tan\beta=7$. \textit{Right}: Branching ratio for each of $\tilde{\chi}^0_2\rightarrow \tilde{\chi}^0_1 h_{1,2,3}$ and $\tilde{\chi}^0_2\rightarrow \tilde{\chi}^0_1 f\bar{f}$, $\tilde{\chi}^0_2\rightarrow \tilde{\chi}^0_1 f\bar{f}$ and $\tilde{\chi}^0_2\rightarrow \tilde{\chi}^0_1 Z$  (the latter three labelled ``Other'') as a function of $m_{\tilde{\chi}^0_2}$, for $M_{h_1}=40\,\mathrm{GeV}$, $\tan\beta=5.5$.}
\end{figure}
In Fig.~\ref{argAtratio1} (left), we show how the percentage effect depends on $A_{t}$.   
The effect is maximised for $\phi_{A_t}=\pi$ where $h_1$ is mostly CP-even 
(and experimentally excluded for this $M_{h_1}$ value).
Hence there can exist 
CP-conserving scenarios in which the effect of our genuine vertex corrections can be even larger than in the CPX scenario.  The dotted curves 
show the reduced effect for
 smaller values of $|A_{t}|$.

For phenomenology it is important to consider branching ratios, which we compute 
using \texttt{FeynArts} and \texttt{FormCalc},
 incorporating our loop-corrected decay widths for $\tilde{\chi}^0_{2}\rightarrow\tilde{\chi}^0_1 h_{1,2,3}$.
The resulting branching ratios of $\tilde{\chi}^0_2$ are plotted against $m_{\tilde{\chi}^0_{2}}$ in Fig.~\ref{argAtratio1} (right). 
For $m_{\tilde{\chi}^0_{2}}\lsim190\,\mathrm{GeV}$, $BR(\tilde{\chi}^0_2\rightarrow\tilde{\chi}^0_1 h_1)\approx 100\%$
and the vertex corrections have negligible effect on the branching ratio.  As we increase $m_{\tilde{\chi}^0_{2}}$, the on-shell decays
$\tilde{\chi}^0_2\rightarrow\tilde{\chi}^0_1 h_{2,3}$
 become kinematically allowed.  The three competing decay widths all receive large 
vertex corrections of order $50\%$, but their 
effects cancel to the order of a few percent for the branching ratios.
For $m_{\tilde{\chi}^0_2}\gsim470$GeV, 
 decays into sfermions become kinematically allowed, and the vertex corrections can alter the branching ratios 
by more than $10\%$. 

Decays of this kind could help to close the ``CPX hole'' at the LHC, using the decay chain;
\begin{equation}
 \widetilde{g} \rightarrow \widetilde{f} \bar{f}\rightarrow \widetilde{\chi}^0_2 f \bar{f} \rightarrow \widetilde{\chi}^0_1 f \bar{f} h_i  \rightarrow \widetilde{\chi}^0_1 f \bar{f} b \bar{b} (\tau^+ \tau^-). 
\end{equation}
 Summing over the 
various decay modes and
combining with $BR(\widetilde{\chi}^0_2\rightarrow\widetilde{\chi}^0_1 h_1)\approx79\%$ for $M_{h_1}\approx 40\, \mathrm{GeV}$, we estimate that around 
$13\%$ of the gluinos produced in this scenario will decay into $h_1$.  This represents a
 large new source of light Higgs bosons.  The question remaining is whether it is possible to dig such
 a signal out of SM and SUSY backgrounds. Existing CMS analyses for an mSUGRA Higgs boson of mass 
$115\,\mathrm{GeV}$ have been promising \cite{Ball:2007zza}.  However, for the CPX hole, the Higgs
 mass is much lighter, so further investigation,
 beyond the scope of this work, would be needed to determine the feasibility of this channel for Higgs discovery.
However, we hope 
these results will encourage further experimental studies.

\begin{theacknowledgments}
\begin{small}Thanks go to my supervisor and collaborator on this work, Georg Weiglein.
This work has been supported
in part by the European Community's Marie-Curie Research
Training Network under contract MRTN-CT-2006-035505
`Tools and Precision Calculations for Physics Discoveries at Colliders'
(HEPTOOLS) and MRTN-CT-2006-035657
`Understanding the Electroweak Symmetry
Breaking and the Origin of Mass using the First Data of ATLAS'
(ARTEMIS).          \end{small}
\end{theacknowledgments}

\end{document}